\documentclass[12pt,preprint]{aastex}

\def\eg{{\it e.g.,\ }}

\slugcomment{Astrophysical Journal}

\shorttitle{H$_2$O Ice Absorption in ULIRGs}
\shortauthors{Imanishi et al.}

\begin{document}

\title{3.1 $\mu$m H$_{2}$O Ice Absorption in LINER-Type 
Ultraluminous Infrared Galaxies with Cool Far-Infrared Colors: the
Centrally-Concentrated Nature of Their Deeply Buried Energy Sources}  

\author{Masatoshi Imanishi\altaffilmark{1,2}}
\affil{National Astronomical Observatory, Mitaka, Tokyo 181-8588, Japan}
\email{imanishi@optik.mtk.nao.ac.jp} 

\and 

\author{Philip R. Maloney}
\affil{Center for Astrophysics and Space Astronomy, 
University of Colorado, Boulder, CO 80309-0839, U.S.A.}
\email{maloney@casa.colorado.edu}
       
\altaffiltext{1}{Visiting Astronomer at the Infrared Telescope Facility,
which is operated by the University of Hawaii under contract from the 
National Aeronautics and Space Administration.}

\altaffiltext{2}{Based in part on data collected at Subaru Telescope,
which is operated by the National Astronomical Observatory of Japan.}

\begin{abstract}

Ground-based 2.8--4.1 $\mu$m slit spectra of the nuclei of seven 
ultraluminous infrared galaxies (ULIRGs) that are classified optically
as LINERs and have cool far-infrared colors are presented. All the
nuclei show 3.3 $\mu$m polycyclic aromatic hydrocarbon (PAH) emission,
with equivalent widths that are systematically lower than those in
starburst galaxies. Strong 3.1 $\mu$m H$_{2}$O ice absorption, with
optical depth greater than 0.6, is also detected in five nuclei, and
3.4 $\mu$m carbonaceous dust absorption is detected clearly in one of the
five nuclei. It is quantitatively demonstrated that the
large optical depths of the H$_{2}$O ice absorption in the five sources,
and the 3.4 $\mu$m absorption in one source, are incompatible with a
geometry in which the energy sources are spatially mixed with dust and
molecular gas, as is expected for a typical starburst, but instead require
that a large amount of nuclear dust (including ice-covered grains) and
molecular gas be distributed in a screen in front of the 3--4 $\mu$m
continuum-emitting sources. This geometrical requirement can naturally be
met if the energy sources are more centrally concentrated than the nuclear
dust and molecular gas. The low equivalent widths of the PAH
emission compared to starbursts and the central concentration of the
nuclear energy sources in these five ULIRGs are best explained by the
presence of energetically important active galactic nuclei deeply
buried in dust and molecular gas.

\end{abstract}

\keywords{galaxies: active --- galaxies: nuclei --- infrared: galaxies
-- galaxies: individual (UGC 5101, IRAS 00188-0856, IRAS 03250+1606, IRAS
16487+5447, IRAS 21329-2346, IRAS 23234+0946, and IRAS 23327+2913)} 

\section{Introduction}

Galaxies that radiate most of their extremely large, quasar-like
luminosities (${^{\displaystyle >}_{\displaystyle \sim}}$
10$^{12}L_{\odot}$) as infrared dust emission -- the
ultraluminous infrared galaxies (ULIRGs; Sanders \& Mirabel 1996) --
dominate the bright end of the galaxy luminosity function in the
nearby universe \citep{soi87}.  They have been used, extensively,
to derive information on the dust-obscured star-formation rate, dust
content, and metallicity in the early universe \citep{bar00,bla99}.
Understanding the nature of nearby ULIRGs, and more particularly,
determining whether they are powered by starbursts or active galactic
nuclei (AGNs), or both, is of great importance in modern
extragalactic astronomy.

Spectroscopic observation of the thermal infrared wavelength range
(3--20 $\mu$m) is currently one of the most powerful methods of
determining the energy sources of ULIRGs.  At this wavelength range,
dust extinction is lower than at shorter wavelengths ($<$2 $\mu$m), so
it becomes possible to detect and measure the emission from more
highly obscured energy sources, with smaller uncertainty in dust
extinction corrections.  In addition, spectral features in this
waveband can be used to distinguish between starburst and AGN
activity.  Polycyclic aromatic hydrocarbon (PAH) emission is detected
in starbursts but not in AGNs, making it a good indicator of starburst
activity \citep{moo86,roc91}, while the presence of AGNs hidden behind
dust can be recognized through absorption features.  
PAH emission and absorption features, detected in ground-based slit spectra
(e.g., Dudley 1999; Imanishi \& Dudley 2000; Soifer et al. 2002) and in
{\it ISO} spectra taken with large apertures (e.g., Tran et al. 2001), have 
been utilized to investigate the energy sources of ULIRGs.

In ULIRGs, there are two potentially energetically significant
components. One is a weakly obscured 
(A$_{\rm V}$ ${^{\displaystyle <}_{\displaystyle \sim}}$ 20 mag) 
\footnote{ Throughout this paper, ``weakly obscured'' is used to mean
that the flux attenuation at 3--4 $\mu$m is insignificant, which is
roughly the criterion that 
A$_{\rm 3-4 \mu m}$ ${^{\displaystyle <}_{\displaystyle \sim}}$ 1 mag, 
or 
A$_{\rm V}$ ${^{\displaystyle <}_{\displaystyle \sim}}$ 20 mag 
if the Galactic extinction curve of 
$A_{\rm 3-4 \mu m}/A_{\rm V} \sim 0.06$ \citep{rie85,lut96} is adopted.}
starburst in the extended (kpc scale) host galaxy and also at the
nucleus. The other is a nuclear, compact (less than a few hundred pc in
size), highly obscured energy source, powered by an AGN and/or a compact
starburst. 
Recent ground-based high spatial resolution imaging observations
\citep{sur99,sur00} have shown that the weakly obscured 
($A_{\rm V} < 20$ mag) starbursts are energetically insignificant, and that
the nuclear, compact, highly obscured energy sources dominate in ULIRGs
\citep{sak99,soi00}. However, the nature of these nuclear sources is
not easy to determine, because dust extinction can be significant even
in the thermal infrared wavelength range.
The observed PAH-to-infrared luminosity ratios in ULIRGs have been
found to be smaller than in known starburst galaxies by a large factor
\citep{fis00}. Assuming that the luminosities of the PAH emission are 
a good indicator of the absolute magnitudes of starbursts, the smaller
ratios in ULIRGs could be due either to an increase in the dust
extinction of the PAH emission, or a dominant contribution from AGNs
to the infrared luminosities of ULIRGs. Determining which explanation
is correct requires reliable estimates of the dust extinction toward
the starbursts in ULIRGs. However, these estimates become complicated
if the assumption of a ``foreground screen'' dust geometry is not
valid for the starbursts \citep{mcl93}.

To understand the nature of the compact, highly obscured energy
sources in ULIRGs, the spectral shapes of the nuclear compact emission
can provide more insight than just the absolute luminosities of PAH
emission, because the equivalent widths of PAH emission are, by
definition, robust to the amount of dust extinction. If the nuclear
compact emission of ULIRGs displayed significantly lower equivalent
widths of PAH emission than typically seen in starburst galaxies, and,
in addition, strong absorption features were detected, it would
suggest that energetically important dust-obscured AGNs were present.
Since the attenuation of the extended, weakly obscured emission is
much smaller than that of the emission from the compact, highly
obscured energy sources, the observed 3--20 $\mu$m spectra taken
with large apertures can be strongly contaminated by the extended
emission, even though the total infrared (8--1000 $\mu$m) luminosities
of ULIRGs are dominated by the compact energy sources.  In this case,
the signatures of absorption features toward the compact energy
sources can be diluted.  Hence spectroscopy with a narrow (less than a
few arcsec) slit -- comparable to the spatial extent of the compact
component -- is better suited to probe the nature of the compact  
energy sources in ULIRGs.

Ground-based slit spectroscopy in the $L$-band (2.8--4.1 $\mu$m) has several
important advantages for investigating the nature of dust-enshrouded
AGNs:  
\begin{enumerate}
\item The dust extinction curve is fairly grey (wavelength-independent) 
at 3--10 $\mu$m \citep{lut96}. 

\item The dust around an AGN has a strong temperature gradient, in
that the inner (outer) dust has a higher (lower) temperature.  The
temperature of the innermost dust is determined by the dust
sublimation temperature (T $\sim$ 1000 K).  In the case of blackbody
radiation ($\lambda_{\rm peak}$ [$\mu$m] $\times$ T [K] $\sim$ 3000),
the 3--4 $\mu$m continuum emission is dominated by dust with $T =
800$--1000 K located in the innermost regions, while the $\sim$10
$\mu$m continuum emission is dominated by dust with $T \sim 300$ K
located further out. Thus, the dust extinction toward the $\sim$10
$\mu$m continuum emission regions, as estimated using \eg the optical
depth of the 9.7 $\mu$m silicate dust absorption feature, is much
smaller than the dust extinction toward the AGN; the dereddened AGN
luminosity will be underestimated if the extinction correction is made
based on $\sim$10 $\mu$m data alone.  By contrast, the dust extinction
toward the innermost, 3--4 $\mu$m continuum emission regions, as
estimated using 3--4 $\mu$m data, is a better indicator of the dust
extinction toward the AGN itself. In the case that an AGN is the only
energy source, the 3--4 $\mu$m continuum emission predominantly
originates in the innermost, 800--1000 K dust. The dust extinction
toward the 3--4 $\mu$m continuum emission can therefore be reliably
fitted with a foreground screen dust model. In a real galaxy, weakly
obscured starburst emission may contribute to the observed
nuclear 3--4 $\mu$m flux. However, as long as the AGN is sufficiently
luminous that AGN-powered continuum emission contributes significantly
to the observed 3--4 $\mu$m flux, evidence for an AGN can be found,
provided that the starburst contribution is subtracted properly, even
if the absorption optical depth toward the AGN is larger than unity at
3--4 $\mu$m ($A_{3-4\mu m} > 1$ mag or $A_{\rm V} > 20$ mag; Rieke \&
Lebofsky 1985). We have previously demonstrated this in two ULIRGs:
IRAS 08572+3915 ($A_{\rm V} > 130$ mag; Imanishi \& Dudley 2000), and
UGC 5101 ($A_{\rm V} > 115$ mag; Imanishi et al. 2001).

\item At 2.8--4.1 $\mu$m, in addition to the 3.3 $\mu$m PAH emission, 
3.4 $\mu$m carbonaceous dust absorption and 3.1 $\mu$m H$_2$O ice
absorption are present. As we will show later in this paper,
all these features can be used in a complementary manner to constrain
the properties of the compact energy sources. They are simultaneously
observable with recently available spectrographs that cover the whole
$L$-band atmospheric window (2.8--4.1 $\mu$m), if the target sources
are at $z < 0.15$.  In this redshift range, a statistically
significant number of ULIRGs have been found with {\it IRAS}.

\item In the near future, slit spectra of ULIRGs at 5.3--40 $\mu$m
will be obtainable using the {\it SIRTF} IRS.  By combining 3--4
$\mu$m and 5.3--40 $\mu$m slit spectra, the presence of a strong
temperature gradient in the dust (a signature of an AGN) can be
investigated through the comparison of the optical depths of
absorption features at different wavelengths at 3--20 $\mu$m
\citep{dud97,ima00}.
\end{enumerate}

In the study of ULIRGs, it is particularly important to detect and
quantify elusive AGNs that are deeply buried in dust along all
sightlines (hereafter buried AGNs; Imanishi et al. 2001). ULIRGs that
are classified optically as LINERs and have cool far-infrared colors
(Sanders et al. 1988) have generally been taken to be starburst-powered
\citep{dow98,tan99,lut99}. However, \citet{idm01} argued that LINER
ULIRGs may contain buried AGNs (see also Antonucci 2001). In this
paper 2.8--4.1 $\mu$m ground-based slit spectra of LINER ULIRGs with
cool far-infrared colors are reported, to search for observational
evidence for energetically important, buried AGNs. Throughout this
paper, H$_{0}$ $=$ 75 km s$^{-1}$ Mpc$^{-1}$, $\Omega_{\rm M}$ = 0.3,
and $\Omega_{\rm \Lambda}$ = 0.7 are adopted.

\section{Targets, Observations and Data Analysis}

The seven LINER ULIRGs in Table~\ref{tbl-1} were observed. 
Table~\ref{tbl-1} summarizes their infrared emission properties. 
All have cool far-infrared colors.  
The optical classifications as LINERs are based on \citet{vei95} and 
\citet{vei99}. 
Table~\ref{tbl-2} summarizes the observing log.

The 3.1--4.0 $\mu$m slit spectrum of UGC 5101 has already been
presented in \citet{idm01}. Descriptions of the observations and data
analysis were offered in that Letter. To summarize, the observations
were made at the IRTF on Mauna Kea, Hawaii, using the grism mode of
NSFCAM \citep{shu94}. Although the wavelength coverage of the NSFCAM
grism mode is 2.8--4.1 $\mu$m, results at $\lambda<$ 3.1 $\mu$m were
not presented in \citet{idm01} because our main aim in that study was
to investigate the 3.3 $\mu$m PAH emission and 3.4 $\mu$m carbonaceous
dust absorption features, and because the Earth's atmospheric
transmission at 2.8--3.1 $\mu$m is poorer than at $\lambda >$ 3.3
$\mu$m.  However, after the publication of \citet{idm01},
\citet{spo02} reported the detection of 6.0 $\mu$m H$_{2}$O ice
absorption in UGC 5101, and suggested that the 3.1 $\mu$m H$_{2}$O ice
absorption feature might be present in our 3.1--4.0 $\mu$m
spectrum. Therefore we carefully analyzed the spectrum at 2.8--3.1
$\mu$m to investigate the 3.1 $\mu$m H$_{2}$O ice absorption feature,
whose profile is very broad, extending over at least $\lambda_{rest}
=$ 2.8--3.3 $\mu$m in the rest-frame (Smith, Sellgren, \& Tokunaga
1989).

The 2.8--4.1 $\mu$m spectra of the other ULIRGs were obtained with the
Infrared Spectrograph and Camera (IRCS; Kobayashi et al. 2000) at the
Subaru telescope on Mauna Kea, Hawaii. The sky was clear during the
observations of these objects and the seeing at K was measured to be
0$\farcs$5--0$\farcs$8 in full-width at half maximum. A 0$\farcs$9-wide
slit and the $L$-grism were used with a 58-mas pixel scale. The achievable 
spectral resolution is $\sim$140 at 3.5 $\mu$m. A standard telescope
nodding technique, with a throw of 7 arcsec along the slit, was
employed to subtract background emission. ULIRGs and corresponding
standard stars were observed with an airmass difference of $<$0.1 to
correct for the transmission of the Earth's atmosphere, and to provide
flux calibration.

Standard data analysis procedures were employed, using IRAF   
\footnote{
IRAF is distributed by the National Optical Astronomy Observatories, 
which are operated by the Association of Universities for Research 
in Astronomy, Inc. (AURA), under cooperative agreement with the 
National Science Foundation.}. 
Initially, bad pixels were replaced with interpolated signals from the
surrounding pixels. Bias was subsequently subtracted from the
obtained frames and the frames were divided by a spectroscopic flat
frame. Finally the spectra of ULIRGs and standard stars were
extracted. Wavelength calibration was performed taking into account
the wavelength-dependent transmission of the Earth's atmosphere.  The
spectra of ULIRGs were divided by the observed spectra of standard
stars, multiplied by the spectra of blackbodies with temperatures
appropriate to individual standard stars (Table~\ref{tbl-2}), and then
flux-calibrated.

\begin{figure}
\plottwo{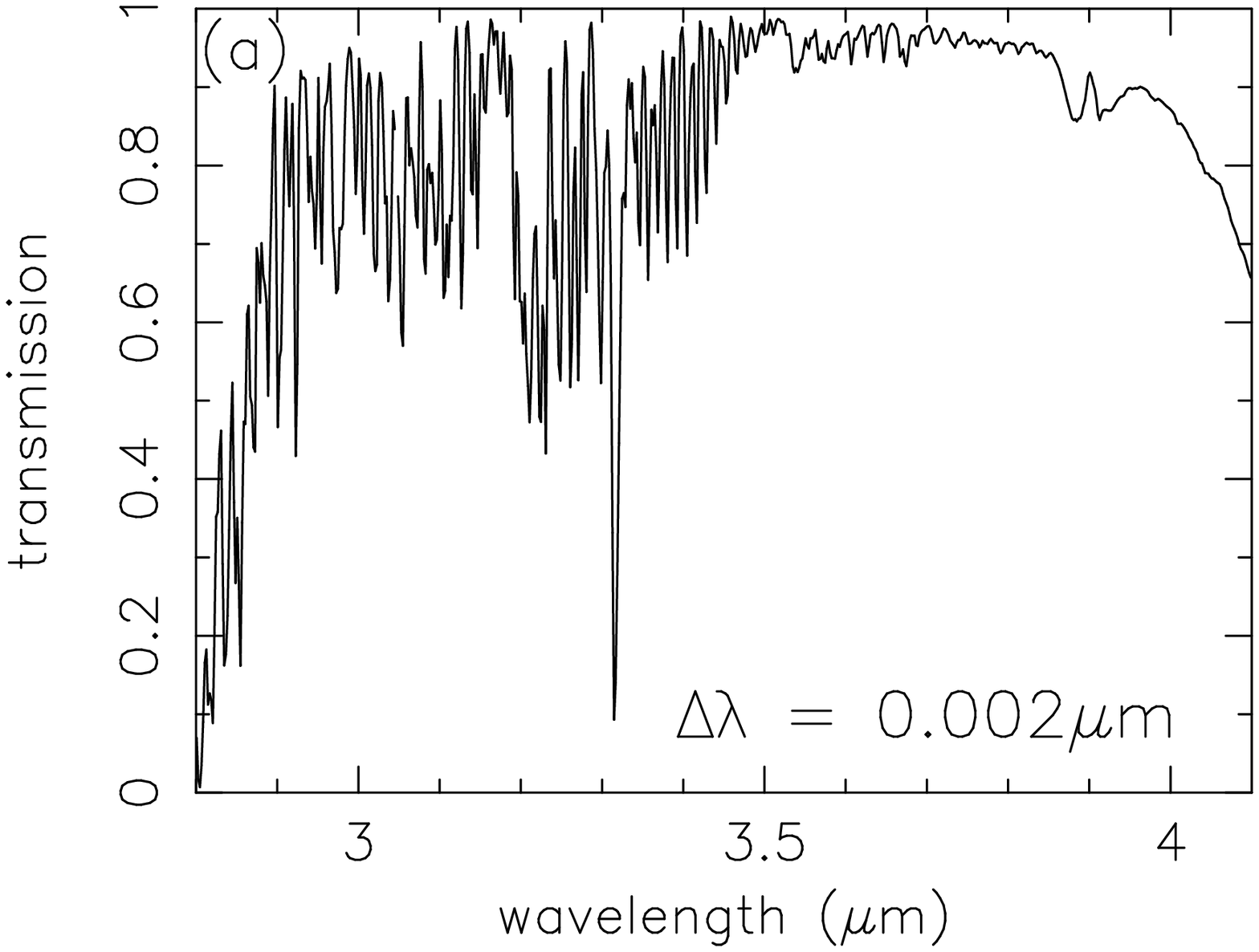}{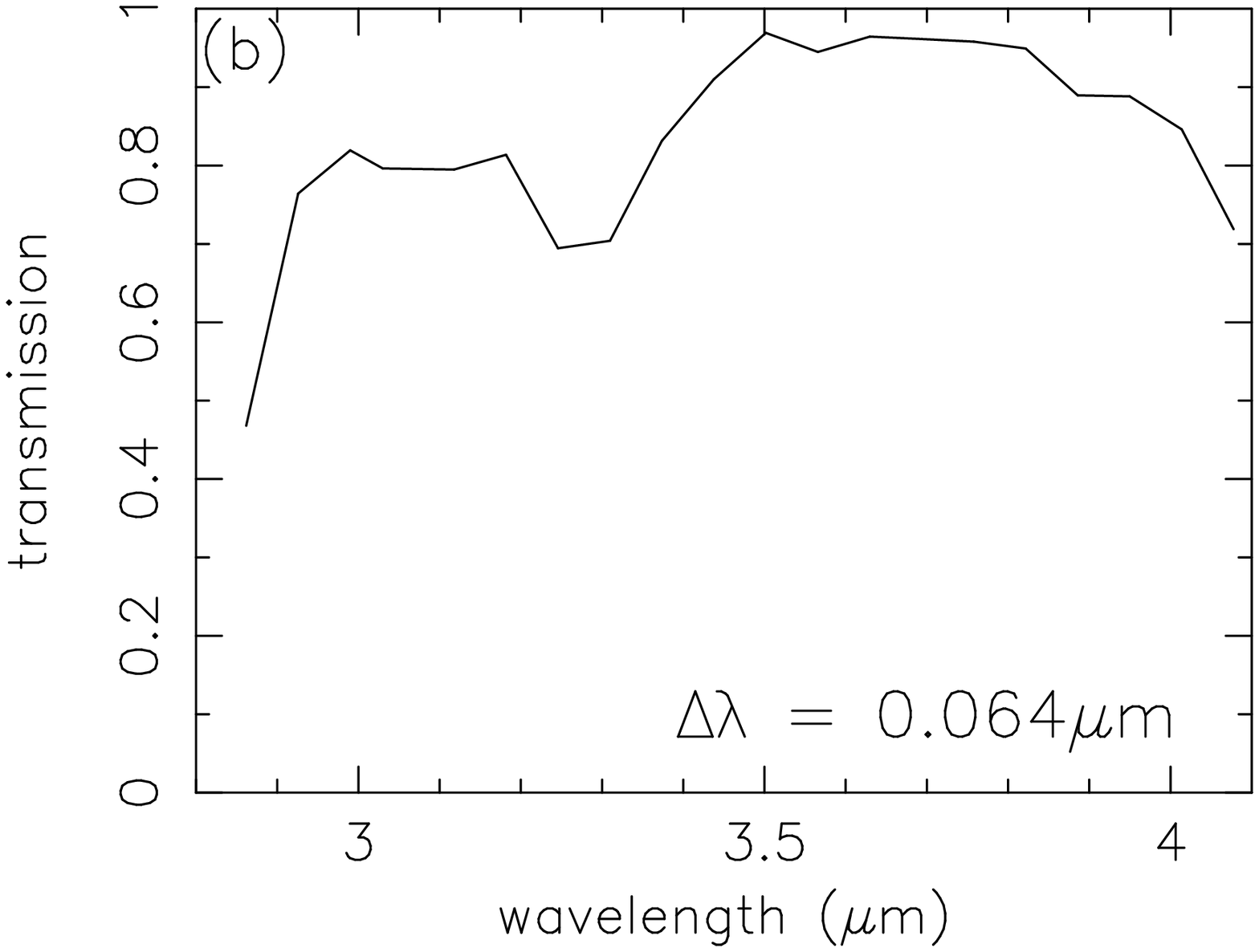}
\caption{
The Earth's atmospheric transmission at 2.8--4.1 $\mu$m at
the summit of Mauna Kea under good weather conditions.  Adopted from
the UKIRT webpage at
http://www.jach.hawaii.edu/JACpublic/UKIRT/astronomy/calib/index1-tx.html.
{\it (a)}: Measured with a spectral resolution of 1500--2000 at 3--4
$\mu$m ($\Delta\lambda =$ 0.002 $\mu$m).  {\it (b)}: Measured with a
spectral resolution of 50--60 at 3--4 $\mu$m ($\Delta\lambda = 0.064$
$\mu$m), roughly the same as the spectral resolution of the seven ULIRGs 
at $\lambda_{observed} < 3.3$ $\mu$m.  Note that the transmission
becomes worse with increasing precipitable water, particularly at
2.8--3.3 $\mu$m.
\label{fig1}}
\end{figure}

Appropriate binning of spectral elements was performed, particularly at
$\lambda_{observed} < 3.3$ $\mu$m in the observed frame, so as to give
an adequate signal-to-noise ratio in each element.  The resulting
spectral resolution $R$ at $\lambda_{observed} < 3.3$ $\mu$m is
${^{\displaystyle <}_{\displaystyle \sim}}$100. 
Although the Earth's atmospheric transmission curve at
2.8--3.3 $\mu$m is highly wavelength-dependent if observed at high
spectral resolution of $R > 1000$ (Figure~\ref{fig1}a), it is fairly
smooth at lower spectral resolutions, with 
R ${^{\displaystyle <}_{\displaystyle \sim}}$ 100
(Figure~\ref{fig1}b). Thus, even if the net positions of the target 
object and standard star on the slit differ slightly (on the sub-pixel
scale) along the wavelength direction, the standard data analysis
described above is expected to produce no significant spurious
features in spectra with 
R ${^{\displaystyle <}_{\displaystyle \sim}}$ 100.

\section{Results}

Figure~\ref{fig2} shows flux-calibrated 2.8--4.1 $\mu$m spectra of the
seven ULIRGs.  
Since our slit spectra cover physical scales larger than a few hundred pc
in all the ULIRGs, emission from both AGNs and compact (less than a few
hundred pc) starbursts should be fully detected.

All the ULIRGs clearly show the 3.3 $\mu$m PAH emission. To estimate
its flux, luminosity, and rest-frame equivalent width, we make the
reasonable assumption that the profiles of the 3.3 $\mu$m PAH emission
in these ULIRGs are similar to those of Galactic star-forming regions
and nearby starburst galaxies; the main emission profile then extends
between $\lambda_{rest}$ $=$ 3.24--3.35 $\mu$m \citep{tok91,imd00}.
The flux excess above the solid lines in Figure~\ref{fig2} should thus be
ascribed to the 3.3 $\mu$m PAH emission. Table~\ref{tbl-3} summarizes the
results for the 3.3 $\mu$m PAH emission in these seven ULIRGs.

The spectrum of UGC 5101 shows clear 3.4 $\mu$m carbonaceous dust
absorption with an observed optical depth of $\tau_{3.4}$ = 0.65
\citep{idm01}. In other ULIRGs, there is no clear indication of this
feature, partly because this absorption feature is intrinsically not so
strong ($\tau_{3.4}$/A$_{\rm V}$ $\sim$ 0.004--0.007; 
Pendleton et al. 1994).   

In addition to the 3.3 $\mu$m PAH emission feature,
there is another important common feature of the spectra of five ULIRGs 
(UGC 5101, IRAS 00188$-$0856, IRAS 03250+1606, IRAS 16487+5447, and 
IRAS 21329$-$2346):  
the continuum emission is concave in these spectra. At the shorter
wavelength side of the 3.3 $\mu$m PAH emission, the continuum flux
level initially decreases with decreasing wavelength, but then begins
to increase again at $\lambda_{rest} \sim 3.05$ $\mu$m.  In IRAS
00188$-$0856 and IRAS 03250+1606, the spectra become flat at the
shortest wavelength parts ($\lambda_{rest}$ $<$ 2.7 $\mu$m). This
behavior is quite similar to that of the infrared-luminous galaxy NGC 4945,
in which 3.1 $\mu$m H$_{2}$O ice absorption was clearly detected in an 
{\it ISO} spectrum \citep{spo00}.

Earlier ground-based $L$-band spectroscopy has revealed that many
bright Galactic sources behind substantial columns of dust and
molecular gas show broad 3.1 $\mu$m H$_{2}$O ice absorption, with a
main profile peaking at $\lambda_{rest}$ $\sim$ 3.05 $\mu$m and
extending at least to $\lambda_{rest}$ $=$ 2.8--3.3 $\mu$m (Smith et
al. 1989; Smith, Sellgren, \& Tokunaga 1993; Smith 1993). A recent
{\it ISO} spectrum of NGC 4945 suggests that the absorption may extend
to wavelengths as short as $\lambda_{rest} \sim 2.5$ $\mu$m
\citep{spo00}. In addition to this main profile, a weak
longer-wavelength wing at $\lambda_{rest} = 3.3$--3.7 $\mu$m can
also be distinguished \citep{smi89,smi93b,smi93}. The concave-shaped
continuum emission found in the spectra of these five ULIRGs is
consistent with the presence of this broad 3.1 $\mu$m H$_{2}$O ice
absorption.

Some Seyfert 2 galaxies with lower infrared luminosities were
observed on the same nights as these ULIRGs, but no signatures of strong
3.1 $\mu$m H$_{2}$O ice absorption were found. The spectra of standard
stars observed before these ULIRGs on the same nights were divided by
other standard stars observed after, but the resulting spectra also
never showed such signatures. The strong 3.1 $\mu$m H$_{2}$O ice
absorption signatures found only in the spectra of the ULIRGs are thus
believed to be genuine and intrinsic to the ULIRGs, and not an
artifact caused by insufficient cancellation of the Earth's
atmospheric transmission. In fact, for UGC 5101 and IRAS 00188$-$0856,
the clear detection of H$_{2}$O ice absorption at 6.0 $\mu$m was also
reported by \citet{spo02}.

To estimate conservative lower limits for the optical depths of the broad 
3.1 $\mu$m H$_{2}$O ice absorption feature, the concave quadratic dashed
lines in Figure~\ref{fig2} are adopted as the lowest plausible continuum
levels.  
In IRAS 23234+0946 and IRAS 23327+2913, the presence of strong 3.1 $\mu$m
H$_{2}$O ice absorption is not clear in the case of the adopted concave
continuum.   
For the remaining five ULIRGs, the observed optical depths of the 3.1
$\mu$m H$_{2}$O absorption ($\tau_{3.1}$) as well as the 3.4 $\mu$m
carbonaceous dust absorption ($\tau_{3.4}$) are summarized in
Table~\ref{tbl-4}.  
These $\tau_{3.1}$ values should be regarded as strict lower limits because:
(1) linear continuum levels that connect data points at the shortest and
    longest wavelength parts of the spectra would give rise to larger
    values of $\tau_{3.1}$, and  
(2) the data points at the shortest wavelength parts, used to determine
    continuum levels, may still be affected by the 3.1 $\mu$m H$_{2}$O ice
    absorption feature in some cases. 
Compared to UGC 5101, IRAS 00188$-$0856 and IRAS 21329$-$2346, the presence
of the broad H$_{2}$O ice absorption feature is apparently less clear in
IRAS 03250+1606 and IRAS 16487+5447. However, as long as the spectral
shapes of the 3.3 $\mu$m PAH emission in these ULIRGs do not differ
significantly from those observed so far in starburst galaxies, the
continuum flux levels on the shorter wavelength side of the 3.3 $\mu$m PAH
emission are clearly lower than on the longer wavelength side, which is
consistent with the presence of the broad 3.1 $\mu$m H$_{2}$O ice
absorption. The very broad emission-like features that extend between
$\lambda_{observed} =$ 3.6--3.9 $\mu$m and 3.4--3.8 $\mu$m in IRAS
03250+1606 and IRAS 16487+5447, respectively, cannot be ascribed
solely to the 3.3 $\mu$m PAH emission; instead, they are caused by a
combination of 3.1 $\mu$m H$_{2}$O ice absorption, 3.3 $\mu$m
PAH emission, and (possibly) 3.4 $\mu$m carbonaceous dust absorption.

\section{Discussion}

\subsection{Energetic importance of weakly obscured starbursts}

The observed 3.3 $\mu$m PAH to infrared luminosity ratios for these
seven ULIRGs are L$_{\rm 3.3PAH}$/L$_{\rm IR}$ $\sim$ 1--3 $\times
$10$^{-4}$ (Table~\ref{tbl-3}), which are 3 to 10 times smaller than
the ratios for starburst galaxies (L$_{\rm 3.3PAH}$/L$_{\rm IR}$
$\sim$ 10$^{-3}$; Mouri et al. 1900; Imanishi 2002). It can therefore
be concluded that {\it weakly obscured (A$_{\rm V} < 20$ {\rm mag})
nuclear} starbursts are not contributing significantly to the infrared
luminosities of these ULIRGs.

Since the L$_{\rm 3.3PAH}$ values are based on slit spectroscopy
measurements, the luminosities of {\it extended} (kpc scale) weakly
obscured starbursts are not taken into account. 
However, as mentioned in $\S$1, such extended starbursts
have been shown to be energetically insignificant in ULIRGs
\citep{sur99,sur00,fis00}. For UGC 5101, \citet{spo02} estimated,
based on the {\it ISO}'s large aperture measurement, a 6.2 $\mu$m PAH
emission flux of $2.0 \times 10^{-19}$ W cm$^{-2}$, so that its luminosity
is $L_{\rm 6.2PAH} \sim 6.5 \times 10^{42}$ ergs s$^{-1}$ in our adopted
cosmology. The resulting $L_{\rm 
6.2PAH}/L_{\rm IR} \sim 1.6 \times 10^{-3}$, which is only $\sim 1/4$
of the ratio typically seen in starburst galaxies ($6\times 10^{-3}$:
Fischer 2000).  Thus, weakly obscured extended starbursts are unlikely
to be the dominant energy source in UGC 5101, either. 
The dominant energy sources should be compact and heavily obscured; for the
five ULIRGs, the presence of such buried energy sources is supported by the
detection of strong absorption features in our 2.8--4.1 $\mu$m spectra.

\subsection{Geometrical requirements on the buried energy sources} 

The detection of strong ($\tau_{3.1}$ $>$ 0.6) 3.1 $\mu$m H$_{2}$O ice
absorption can, in principle, provide stringent constraints on the
nature of the buried energy sources at the nuclei of the five ULIRGs. The
3.4 $\mu$m carbonaceous and 9.7 $\mu$m silicate dust 
absorption features are detectable if an adequate amount of dust in
the {\it diffuse inter-stellar medium} is present in front of the
continuum-emitting energy sources. However, in order for H$_{2}$O ice
absorption to be detected, dust grains {\it covered with an ice
mantle} must be present in the foreground. Such a situation can arise
only when the dust grains are located in molecular clouds, and are
sufficiently shielded from the UV radiation from any energy sources
either inside or outside the molecular clouds \citep{whi88}.  In other
words, the detection of strong H$_{2}$O ice absorption requires that a
large amount of dust in {\it molecular clouds} be present in front of
the continuum-emitting energy sources. Foreground molecular clouds in
the host galaxies could be responsible for the strong 3.1 $\mu$m
H$_{2}$O ice absorption detected toward the nuclei if the host
galaxies were seen from nearly edge-on. However, there is no strong
evidence that the host galaxies of these ULIRGs are edge-on
\citep{mur96,sco00,kim02}. \citet{spo02} reported that 6.0 $\mu$m
H$_{2}$O ice absorption is systematically stronger in ULIRGs than in
less infrared-luminous Seyfert 2 galaxies. These Seyfert 2 galaxies
also generally show no detectable or weak ($\tau_{3.1} < 0.2$) 3.1
$\mu$m H$_{2}$O ice absorption (Imanishi et al. 2002, in preparation).
The systematically larger optical depths of H$_{2}$O ice absorption
toward the nuclei of ULIRGs can reasonably be explained by a larger
amount of highly concentrated molecular gas (and dust) in their nuclei
\citep{sam96}. We thus regard it as likely that the strong
($\tau_{3.1}$ $>$ 0.6) 3.1 $\mu$m H$_{2}$O ice absorption detected
toward the nuclei of these five ULIRGs are caused by nuclear
concentrations of molecular gas and dust. Note that, although the host
galaxy of NGC 4945 is viewed edge-on \citep{abl87}, a significant
fraction of the strong 3.1 $\mu$m H$_{2}$O ice absorption ($\tau_{3.1}
\sim 0.9$; Spoon et al. 2000) is likely to originate in a highly
concentrated nuclear component of ice-covered dust grains, on a scale
of less than a few hundred pc \citep{bro88}.

In the case of {\it normal} starbursts, it seems reasonable to assume
that the continuum-emitting sources and molecular gas and dust are
spatially well mixed \citep{soi00}. A detailed infrared study of the
nearby starburst galaxy M82 suggests that the mixed-dust geometry is a
better fit to the observed data than a foreground screen dust
distribution \citep{mcl93}. Since the oscillator strengths of species
are fixed, there are upper limits on the absorption optical depths
that can be produced by the mixed-dust geometry (unless unusual
abundances are considered). This is because the foreground,
less-obscured, and therefore less-attenuated, emission (which shows only
weak absorption features) will dominate the observed fluxes.
In practice, nearby starburst galaxies such as M82 and NGC 253 show
weak ($\tau$ $\sim$ 0.2) or undetectable absorption by H$_{2}$O ice
\citep{stu00} or dust grains \citep{roc91,duw99,imd00}.  This issue
will be discussed quantitatively below.

In Galactic molecular clouds, the optical depth of the H$_{2}$O ice
absorption and dust extinction ($A_{\rm V}$) are correlated, with a
relation given by $\tau_{3.1} =\alpha \times (A_{\rm V} - A_{\rm
V0})$.  The value of $\alpha$ has been found to be almost constant, 
$\sim$0.06, in Galactic molecular clouds, which exhibit various different
levels of star-forming activity \citep{smi93b,tan90,mur00}.  This is
because it reflects the abundance of H$_{2}$O ice relative to dust
grains in the regions where an ice mantle can survive; it is therefore
unlikely to differ substantially from the Galactic value, even in
starburst galaxies.  The A$_{\rm V0}$ value in the above
relation is the threshold required for dust grains to be sufficiently
shielded from the ambient UV radiation.  This value can vary
dramatically, from 2--6 mag in quiescent molecular clouds (Taurus and
Serpens; Whittet et al. 1988; Eiroa \& Hodapp 1989; Murakawa et
al. 2000) to 15 mag in active star-forming molecular clouds
($\rho$-Oph; Tanaka et al. 1990).  In starburst galaxies with a more
intense UV radiation field than the $\rho$-Oph molecular clouds,
$A_{\rm V0}$ is almost certainly larger than 15 mag, but its exact
value is unknown.  Thus the relation 
\begin{eqnarray}
\tau_{3.1} & = & 0.06 \times (A_{\rm V} - A_{\rm V0}),
\end{eqnarray}
is adopted in starburst galaxies, where $A_{\rm VO} > 15$ mag, 
or 
\begin{eqnarray}
\tau_{3.1} & = & 0.06 \times f \times A_{\rm V}, 
\end{eqnarray}
where $f$ is the fraction of dust that is covered with an ice mantle. 
   
Figures~\ref{fig3}a and ~\ref{fig3}b show schematic diagrams of a
mixed-dust geometry. In this geometry, the observed flux 
I($\tau_{\nu}$) is given by
\begin{eqnarray}
I(\tau_{\nu}) & = & I_{0} \times \frac{1 - e^{-\tau_{\nu}}}{\tau_{\nu}}, 
\end{eqnarray}
where $I_{0}$ is an unattenuated intrinsic flux and $\tau_{\nu}$ is the
optical depth at each wavelength, which takes different values inside
and outside the absorption features.  
For 3--4 $\mu$m continuum emission outside the 3.1 $\mu$m H$_{2}$O ice and
3.4 $\mu$m carbonaceous dust absorption features, 
\begin{eqnarray}
\tau_{cont} & = & 0.06 \times A_{\rm V} 
\end{eqnarray}
\citep{rie85,lut96}, where, by definition, $A_{\rm V} \equiv 1.08 \times
\tau_{V}$. 
At the peak wavelength of the 3.1 $\mu$m H$_{2}$O ice absorption feature, 
\begin{eqnarray}
\tau_{ice} & = & 0.06 \times A_{\rm V} + 0.06 \times f \times A_{\rm V}, 
\end{eqnarray}
where the first and second terms give the flux attenuation for 3--4 $\mu$m 
continuum and the H$_{2}$O ice absorption feature, respectively.
Thus, the optical depth of the H$_{2}$O ice absorption feature in an
observed 3--4 $\mu$m flux is
\begin{eqnarray}
\tau_{3.1} & \equiv & \ln [\frac{1 - e^{-\tau_{cont}}}{\tau_{cont}} 
\times \frac{\tau_{ice}}{1 - e^{-\tau_{ice}}}] \\
& = & \ln [(1+f) \frac{1 - e^{-0.06 \times A_{\rm V}}}{1 - e^{-0.06 \times
A_{\rm V} \times (1+f)}}]  
\end{eqnarray}
This last formula implies that, for a fixed $f$-value, $\tau_{3.1}$
increases with increasing A$_{\rm V}$, but saturates at a certain
value. The $f$-value can be constrained based on observations of the
nearby starburst galaxy M82.  Based on a mixed-dust model, the dust
extinction toward the farthest parts of the starburst region is
estimated to be A$_{\rm V}$ $\sim$ 55 mag \citep{mcl93}.  The observed
optical depth of the 3.1 $\mu$m H$_{2}$O ice absorption is $\tau_{3.1}
\sim 0.2$ \citep{stu00}.  Based on these values, $f \sim 0.3$ is
obtained.  In other words, in the starburst galaxy M82, $\sim30\%$ of
dust grains are covered with an ice mantle.

If the cores of ULIRGs are simply a scaled-up version of normal
starbursts with a mixed-dust geometry and the $f$-value is $\sim$0.3,
then $\tau_{3.1}$ should be $\sim$0.2.  However, the $f$-value is
strongly dependent on the fraction of active star-forming regions
relative to quiescent molecular gas in molecular clouds.  If the cores
of ULIRGs are powered by starbursts with a mixed-dust geometry, the
$f$-value in ULIRGs should be even smaller than in less infrared
luminous starburst galaxies \citep{soi00}.  Here, however, $f$ $\sim$
0.3 is adopted for ULIRGs in order to determine a conservative upper
limit for $\tau_{3.1}$ within the framework of a mixed-dust geometry.
Even if this conservative value is adopted, Equation (7) shows that
$\tau_{3.1}$ cannot be larger than 0.3. Therefore, it can
be concluded that {\it the observed $\tau_{3.1}$ values of $>$0.6 in
these five ULIRGs are clearly incompatible with mixed-dust
geometry and f $ {^{\displaystyle <}_{\displaystyle \sim}}$ 0.3}.
When the contribution from any weakly obscured starburst emission to
the observed 3--4 $\mu$m fluxes is subtracted, the $\tau_{3.1}$ values
toward the buried energy sources will be even larger, making this
conclusion more robust.

The same argument is applicable to the 3.4 $\mu$m carbonaceous dust
absorption in UGC 5101 \citep{idm01}.
For the 3.4 $\mu$m carbonaceous dust absorption feature, 
\begin{eqnarray}
\tau_{carbo} & = & 0.06 \times A_{\rm V} + (0.004\sim0.007) \times (1-f)
\times A_{\rm V}  
\end{eqnarray}
is obtained \citep{pen94}, where the term $(1-f)$, the fraction of dust
grains not covered by an ice mantle, is included because the 3.4 $\mu$m
carbonaceous dust absorption feature may be suppressed if the dust grains are
covered with an ice mantle \citep{men01}.  
Based on the same arguments as above, it can be quantitatively demonstrated
that $\tau_{3.4}$ cannot be larger than 0.2 in the mixed-dust geometry. 
The observed large optical depth of $\tau_{3.4} > 0.6$ in UGC 5101
\citep{idm01} is again incompatible with a mixed-dust source geometry. 

In order to explain the strong absorption features in these five ULIRGs,
dust in a foreground screen must be obscuring the 
3--4 $\mu$m continuum-emitting energy sources. 
This configuration requires that the energy sources be
centrally concentrated and deeply buried in nuclear dust and molecular gas,
as shown in Figure~\ref{fig3}c.  

\subsection{What are the centrally-concentrated energy sources?} 

\subsubsection{The case for UGC 5101} 

The centrally-concentrated nature of the buried energy source in UGC 5101
is naturally explained by a buried AGN \citep{soi00}.
Furthermore, as mentioned in $\S$1, the observed rest-frame equivalent
widths of the 3.3 $\mu$m PAH emission in starburst galaxies
should be little changed by dust extinction.   
The significantly lower equivalent width in UGC 5101 (Table~\ref{tbl-3})
than in starbursts ($\sim$120 nm; Moorwood 1986; Imanishi \& Dudley 2000)
also supports the presence of AGNs that produce strong 
3--4 $\mu$m continuum emission, but essentially no PAH emission. 

When our data are combined with ground-based high spatial resolution
mid-infrared (8--25 $\mu$m) imaging data \citep{soi00}, 
we can give more insight into the nature of the buried energy source. 
\citet{soi00} estimated the size of the
spatially-unresolved core of UGC 5101 at 12.5 $\mu$m to be $<$200 pc.
If the attenuation of the mid-infrared flux from the
spatially-unresolved compact core in UGC 5101 is negligible, the
surface brightness of the core emission becomes $>$1.0 $\times$ 10$^{12}
L_{\odot}$ kpc$^{-2}$ \citep{soi00}.  However, the detection of strong
3.4 $\mu$m carbonaceous dust and 3.1 $\mu$m H$_{2}$O ice absorption
toward the nucleus of UGC 5101 requires that
\begin{enumerate}
\item the actual size of the buried centrally-concentrated energy source
is substantially smaller than that of the dust and molecular gas, and 
\item the dust extinction toward the buried energy source is $A_{\rm V} >
115$ mag \citep{idm01}.  
\end{enumerate}

On point (1), since dust around a centrally-concentrated energy source
shows a temperature gradient (see $\S$1), the size of the cooler dust
region, which is also responsible for the foreground dust extinction,
is larger than that of the $\sim$250 K dust traced by the 12.5 $\mu$m
emission ($<$200 pc). However, the size of the cool dust emission
(traced by the 1.3 mm emission) at the core of the ULIRG Arp 220 has
been shown to be compact ($<$200 pc; Sakamoto et
al. 1999). Furthermore, the sizes of the high density molecular gas
regions at the cores of ULIRGs have been estimated to be no more than
$\sim$500 pc in diameter \citep{sol92}.  Thus, it seems reasonable to
assume that the actual size of the buried centrally-concentrated
energy source in UGC 5101 is only a fraction of this, $\sim$100 pc or
less. On point (2), the dust extinction of $A_{\rm V} > 115$ mag means
that the flux attenuation at 12.5 $\mu$m is more than factor of 25
\citep{rie85}. When these corrections are taken into account, the
actual surface brightness of the centrally-concentrated buried energy
source in UGC 5101 is $> 10^{14}L_{\odot}$ kpc$^{-2}$. This exceeds
the peak surface brightness of star clusters in starburst galaxies
($\sim$5 $\times$ 10$^{13}L_{\odot}$ kpc$^{-2}$; Meurer et al. 1997;
Soifer et al. 2000) by a factor of more than two. The high surface
brightness is best explained by the presence of an energetically
important buried AGN in UGC 5101.  

If a buried AGN is present at the nucleus of UGC 5101, the dust is
expected to show a strong temperature gradient (see $\S$1).  Are the
available data consistent with its presence?  \citet{soi00} could not
explicitly find the signature of a temperature gradient based on their
imaging data of UGC 5101. Here, it is searched for using spectroscopic
data. As mentioned in $\S$1 and \citet{ima00}, if a strong temperature
gradient is present in the dust, the optical depth of the 3.4 $\mu$m
carbonaceous dust absorption feature probes deeper into the dust
distribution than that of the 9.7 $\mu$m silicate dust absorption.
Thus, the $\tau_{3.4}$/$\tau_{9.7}$ ratio toward the nucleus of UGC
5101 is expected to be significantly larger than the value seen when
no significant temperature gradient is present
($\tau_{3.4}$/$\tau_{9.7} \sim 0.06$--0.07; Roche \& Aitken 1984,
1985; Pendleton et al. 1994).

After subtracting the contribution from the weakly obscured starburst
emission to the observed 3--4 $\mu$m flux, the $\tau_{3.4}$ value toward
the buried AGN at the core of UGC 5101 was estimated to be
$\sim$0.8 \citep{idm01}, while only a lower limit for the observed 
$\tau_{9.7}$ value ($>$1.5) was given by \citet{spo02} 
based on a measurement with a large aperture using {\it ISO}.
Although the $\tau_{9.7}$ value is a lower limit, its actual value is
unlikely to exceed 2.0, judging from the actual spectrum of UGC 5101 
(Spoon et al. 2002; their Figure 8).  
Thus the $\tau_{3.4}$/$\tau_{9.7}$ ratio is $> 0.4$. 
The actual $\tau_{9.7}$ value toward the buried AGN in UGC 5101 
could be larger than the observed value, because weakly obscured starburst 
emission is likely to contribute significantly to the observed 
$\sim$10 $\mu$m {\it ISO} flux. 
However, since the equivalent width of the 6.2 $\mu$m PAH emission 
in the {\it ISO} spectrum of UGC 5101 is less than half that of starburst
galaxies \citep{spo02}, at least half of the observed continuum flux at
$>$5 $\mu$m should originate in the buried AGN, implying that the
$\tau_{3.4}$/$\tau_{9.7}$ ratio toward the buried AGN is $>0.2$.     
Thus, these spectroscopic data are consistent with the presence of a
temperature gradient of dust at the core of UGC 5101.   

If the dust around an AGN has a torus-like geometry, low-density
clumpy clouds which are directly illuminated by the central AGN can
exist at relatively large distances from the AGN along the axis of the
torus-geometry dust (the so-called the narrow line regions or
NLRs). NLRs produce the strong high-excitation forbidden emission
lines seen in AGN. However, in a buried AGN, X-ray dissociation
regions (Maloney et al. 1996), rather than NLRs, are expected to
be produced around the AGN, so that high-excitation forbidden emission
lines will be weak, consistent with the observed data of UGC
5101 \citep{gen98}.  

\citet{sol92} and \citet{sol01} found that the infrared to HCN (J=1-0)
luminosity ratios in ULIRGs, including UGC 5101 \citep{gao02}, roughly
follow the relation established in less infrared-luminous starburst galaxies.  
By regarding the HCN (J=1-0) luminosities as a tracer of the mass of high
density ($>$10$^{4}$ cm$^{-3}$) molecular gas where star-formation actually
occurs, \citet{sol92} argued that the infrared luminosities of ULIRGs are
powered by starbursts.  
However, in the luminous inner regions (central few hundred pc) of ULIRGs,
the gas pressures - and therefore molecular gas densities - are inevitably
extremely high, whether the energy sources are starbursts or AGNs 
\citep{bar97,mal99}. 
The relative constancy of the infrared to HCN luminosity ratio from
less infrared luminous starburst galaxies to ULIRGs simply indicates
that much of the infrared emission arises from dust in dense molecular
clouds, which is inevitable whatever the sources of the luminosities of
ULIRGs. 

\subsubsection{The other ULIRGs}

For the two ULIRGs which do not show strong 3.1 $\mu$m H$_{2}$O ice
absorption (IRAS 23234+0946 and IRAS 23327+2913), there is no strong
observational evidence for the presence of buried AGNs. 
However, for the other four ULIRGs which show strong 3.1 $\mu$m H$_{2}$O
absorption (IRAS 00188$-$0856, IRAS 03250+1606, IRAS 16487+5447, and IRAS
21329$-$2346), the centrally-concentrated nature of their buried energy
sources and systematically (roughly a factor of two) lower-equivalent-width
PAH emission than starbursts (Table 3) are also naturally explained by the
presence of energetically important, buried AGNs. 
Assuming that approximately half of the observed 3--4 $\mu$m
fluxes originate in buried AGNs, the $\tau_{3.1}$ and $\tau_{3.4}$
toward the buried AGNs are twice as large as the observed values.
These $\tau_{3.1}$ and $\tau_{3.4}$ values reflect the column density
of foreground dust, toward the buried AGNs, covered with and without
an ice mantle, respectively ($\S$4.2).  Based on the relation of
$\tau_{3.1}$/A$_{\rm V}$ $\sim$ 0.06, $\tau_{3.4}$/A$_{\rm V}$ $\sim$
0.004--0.007 ($\S$ 4.2), and $A_{\rm 3-4\mu m}/A_{\rm V}$ $\sim$ 0.06
($\S$ 1), and on a foreground screen dust model ($\S$ 1), it is found
that the dereddened 3--4 $\mu$m dust emission luminosities powered by
the AGNs could be comparable to the infrared luminosities of these
four ULIRGs.

It is sometimes argued that ULIRGs with cool far-infrared colors
cannot be powered by AGNs because AGNs show warm far-infrared colors (e.g.,
Downes \& Solomon 1998). 
Indeed, many known Seyfert 2 galaxies show warm far-infrared colors
\citep{deg87}. 
However, as mentioned in $\S$4.2, the systematically higher detection rate
of H$_{2}$O ice absorption in ULIRGs, as compared to less
infrared luminous Seyfert 2 galaxies, is naturally explained by a higher
column density of nuclear dust in ULIRGs. 
This explanation is supported by imaging data at $\sim$10 $\mu$m: while
emission from the nuclei of less infrared luminous Seyfert 2 galaxies 
is argued to be optically thin at $\sim$10 $\mu$m \citep{alo01}, that
from the nuclei of ULIRGs is estimated to be optically thick \citep{soi02}.
The higher column densities of dust around buried AGNs in ULIRGs
can naturally produce infrared spectral energy distribution
with cooler color than in less infrared luminous Seyfert 2 galaxies. It is
quite possible that many other ULIRGs with cool far-infrared colors contain
energetically important AGNs deeply buried in dust. 

\citet{tak02} have found that a 60 $\mu$m luminosity function in the local 
universe, derived based on $\sim$15000 galaxies, shows a bump at 
$L_{\rm 60 \mu m} \sim 10^{12}L_{\odot}$, and suggested that AGN activity
begins to be energetically important at 
$L_{\rm 60 \mu m} {^{\displaystyle >}_{\displaystyle \sim}} 
10^{12} L_{\odot}$. 
This result also favors the buried AGN scenario in the majority of nearby
ULIRGs. 

\section{Summary}

Strong ($\tau_{3.1} > 0.6$) 3.1 $\mu$m H$_{2}$O ice absorption has been
detected toward the nuclei of five ULIRGs: UGC 5101, IRAS 00188$-$0856,
IRAS 03250+1606, IRAS 16487+5447, and IRAS 21329$-$2346.   
All these ULIRGs have cool far-infrared colors and are classified optically
as LINERs.    
This detection requires a foreground screen dust geometry, which is best
explained if centrally-concentrated, compact energy sources are
buried in nuclear dust and molecular gas. 
For all the ULIRGs, the observed equivalent widths of the 3.3 $\mu$m PAH
emission are significantly lower than those of starburst galaxies.  
These observational data are naturally explained by the presence of AGNs
deeply buried in dust, and support our previous arguments that LINER ULIRGs
with cool far-infrared colors may possess energetically important, buried
AGNs.  

\acknowledgments

We are grateful to Drs. J. Rayner, B. Golisch, H. Terada, K. Aoki,
B. Potter, D. Scarla, and S. Harasawa for their support during our IRTF and
Subaru observing runs.  
We also thank Dr. C. C. Dudley for his useful comments, and Dr. Y. Gao for
personally showing his preprint prior to publication.
PRM is supported by the NSF under grant AST-9900871.
This research made use of the NASA/IPAC Extragalactic Database
(NED), which is operated by the Jet Propulsion Laboratory, California
Institute of Technology, under contract with the National Aeronautics
and Space Administration.

\clearpage

\begin{deluxetable}{lcrrrrccl}
\tabletypesize{\small}
\tablecaption{Observed LINER ULIRGs and their far-infrared emission
properties.
\label{tbl-1}}
\tablewidth{0pt}
\tablehead{
\colhead{Object} & \colhead{Redshift}   & 
\colhead{f$_{\rm 12}$ \tablenotemark{a}}   & 
\colhead{f$_{\rm 25}$ \tablenotemark{a}}   & 
\colhead{f$_{\rm 60}$ \tablenotemark{a}}   & 
\colhead{f$_{\rm 100}$ \tablenotemark{a}}  & 
\colhead{log L$_{\rm FIR}$ \tablenotemark{b}} & 
\colhead{log L$_{\rm IR}$ \tablenotemark{c}} & 
\colhead{f$_{25}$/f$_{60}$ \tablenotemark{d}}  \\
\colhead{} & \colhead{}   & \colhead{(Jy)} & \colhead{(Jy)} 
& \colhead{(Jy)} & \colhead{(Jy)}  & \colhead{(ergs s$^{-1}$)} &
\colhead{(ergs s$^{-1}$)} & \colhead{} 
}
\startdata
UGC 5101 & 0.040 & 0.25 & 1.03 & 11.54 & 20.23 & 45.5 & 45.5 & 0.09 (cool)\\
IRAS 00188$-$0856 & 0.128 & $<$0.12 & 0.37 & 2.59 & 3.40 & 45.8 &
45.9--46.0 & 0.14 (cool)\\ 
IRAS 03250+1606 & 0.129 & $<$0.10 & $<$0.16 & 1.38 & 1.77 & 45.6 &
45.6--45.7 & $<$0.12 (cool)\\ 
IRAS 16487+5447 & 0.104 & $<$0.08 & 0.20 & 2.88 & 3.07 & 45.6 & 45.7 & 0.07
(cool)\\ 
IRAS 21329$-$2346 & 0.125 & $<$0.08 & $<$0.16 & 1.65 & 2.22 & 45.6 &
45.6--45.7 & $<$0.10 (cool)\\ 
IRAS 23234+0946 & 0.128 & $<$0.06 & $<$0.20 & 1.56 & 2.11 & 45.6 &
45.6--45.7 & $<$0.13 (cool)\\ 
IRAS 23327+2913 & 0.107 & $<$0.06 & 0.22 & 2.10 & 2.81 & 45.6 & 45.6--45.7 
& 0.10 (cool)\\ 
\enddata

\tablenotetext{a}{f$_{12}$, f$_{25}$, f$_{60}$, and f$_{100}$ are 
{\it IRAS} FSC fluxes at 12 $\mu$m, 25 $\mu$m, 60 $\mu$m, and 100 $\mu$m,
respectively.} 

\tablenotetext{b}{Logarithm of far-infrared (40--500 $\mu$m) luminosity 
            in ergs s$^{-1}$ calculated with
            $L_{\rm FIR} = 1.4 \times 1.5 \times 10^{39} \times$ 
            $D$(Mpc)$^{2}$ $\times (2.58 \times f_{60} + f_{100}$) 
            ergs s$^{-1}$ \citep{sam96}.}

\tablenotetext{c}{Logarithm of infrared (8$-$1000 $\mu$m) luminosity
       in ergs s$^{-1}$ calculated with
       $L_{\rm IR} = 2.1 \times 10^{39} \times$ D(Mpc)$^{2}$
       $\times$ (13.48 $\times$ $f_{12}$ + 5.16 $\times$ $f_{25}$ +
       $2.58 \times f_{60} + f_{100}$) ergs s$^{-1}$
       \citep{sam96}.}

\tablenotetext{d}{f$_{25}$/f$_{60}$ $<$ ($>$) 0.2 are called cool
(warm) \citep{san88}.}

\end{deluxetable}

\begin{deluxetable}{llcclccc}
\tabletypesize{\small}
\tablecaption{Observing log \label{tbl-2}}
\tablewidth{0pt}
\tablehead{
\colhead{Object} & 
\colhead{Date} & 
\colhead{Integration} & 
\colhead{P.A. \tablenotemark{a}}  & 
\multicolumn{4}{c}{Standard Stars} \\
\colhead{} & 
\colhead{(UT)} & 
\colhead{(Min)} &
\colhead{($^{\circ}$)} &
\colhead{Name} &  
\colhead{$L$-mag} &  
\colhead{Type} &  
\colhead{$T_{\rm eff}$ (K)}  
}
\startdata
UGC 5101 & 2001 April 9 & 18 & 90 & HR 4112 & 3.4 & F8V & 6200 \\ 
IRAS 00188$-$0856 & 2002 August 19 & 25 & 90 \tablenotemark{b} & HR 72 &
5.0 & G0V & 5930 \\ 
IRAS 03250+1606   & 2002 August 19 & 23.3 & 0 & HR 763 & 4.3 & F7V & 6240 \\
IRAS 16487+5447   & 2002 March 28  & 36 & 70 \tablenotemark{c}& HR 5949 &
6.3 & A0V & 9480\\ 
IRAS 21329$-$2346 & 2002 October 24 & 32 & 90 & HR 7898 &
4.5 & G8V & 5400\\ 
IRAS 23234+0946 & 2002 October 24 & 32 & 120 \tablenotemark{d}& HR 8653 &
4.6 & G8IV & 5400\\ 
IRAS 23327+2913 \tablenotemark{e} & 2002 October 24 & 24 & 90 & HR 8955 &
5.1 & F6V & 6400\\  
\enddata

\tablenotetext{a}{Position angle of the slit.
0$^{\circ}$ corresponds to the north-south direction.
Position angle increases with counter-clockwise on the sky plane.}
\tablenotetext{b}{To avoid the contamination from the emission of the
southern foreground star (Kim, Veilleux, \& Sanders 2002).}
\tablenotetext{c}{To attempt to detect signals from both components of the
double nucleus of this object, which are separated by $\sim$3 arcsec in the
$K$-band (2.2 $\mu$m) image \citep{mur96}.   
Only the western primary nucleus \citep{mur01} was clearly detected in our
2.8--4.1 $\mu$m spectrum.}  
\tablenotetext{d}{To attempt to detect signals from both components of the
double nucleus of this object, which are separated by $\sim$5 arcsec in the
$K$-band image \citep{kim02}.   
Only the western nucleus was clearly detected.}
\tablenotetext{e}{The southern primary nucleus \citep{tru01} was observed.}

\end{deluxetable}

\clearpage

\begin{deluxetable}{lcccc}
\tabletypesize{\small}
\tablecaption{The properties of the 3.3 $\mu$m PAH emission. 
\label{tbl-3}}
\tablewidth{0pt}
\tablehead{
\colhead{(1)} & \colhead{(2)} & \colhead{(3)} & \colhead{(4)} & 
\colhead{(5)}  \\
\colhead{Object} & 
\colhead{f$_{3.3 \rm PAH}$} & 
\colhead{L$_{3.3 \rm PAH}$}  & 
\colhead{L$_{3.3 \rm PAH}$/L$_{\rm IR}$} & 
\colhead{rest EW$_{3.3 \rm PAH}$} \\
\colhead{} & 
\colhead{($\times$10$^{-14}$ ergs s$^{-1}$ cm$^{-2}$)} & 
\colhead{($\times$10$^{41}$ergs s$^{-1}$)}  & 
\colhead{($\times$10$^{-4}$)} & 
\colhead{(nm)}  
}
\startdata
UGC 5101 & 16$\pm$3 & 5.2$\pm$1.0 & $\sim$1 & $\sim$35 \\
IRAS 00188$-$0856 & 3.6$\pm$0.4 & 13$\pm$1 & $\sim$1 & $\sim$50 \\
IRAS 03250+1606 & 3.2$\pm$0.4 & 12$\pm$1 & $\sim$3 & $\sim$80 \\
IRAS 16487+5447 & 2.6$\pm$0.4 & 6.2$\pm$1.0 & $\sim$1 & $\sim$70 \\
IRAS 21329$-$2346 & 1.2$\pm$0.2 & 4.2$\pm$0.8 & $\sim$1 & $\sim$50 \\
IRAS 23234+0946 & 2.6$\pm$0.3 & 9.9$\pm$1.0 & $\sim$2 & $\sim$70 \\
IRAS 23327+2913 & 1.9$\pm$0.3 & 4.8$\pm$0.8 & $\sim$1 & $\sim$50 \\
\enddata

Note. --- 
Col. (1): Object name. 
Col. (2): Observed flux of the 3.3 $\mu$m PAH emission. 
Col. (3): Observed luminosity of the 3.3 $\mu$m PAH emission.  
Col. (4): Observed 3.3 $\mu$m PAH-to-infrared luminosity ratio. 
Col. (5): Rest-frame equivalent width of the 3.3 $\mu$m PAH emission.  
          Those for starbursts are $\sim$120 nm (see text).  

\end{deluxetable}


\begin{deluxetable}{lcc}
\tablecaption{The observed optical depths of the 3.1 $\mu$m H$_{2}$O ice
and 3.4 $\mu$m carbonaceous dust absorption features for the five ULIRGs 
with strong 3.1 $\mu$m H$_{2}$O ice absorption. 
\label{tbl-4}}
\tablewidth{0pt}
\tablehead{
\colhead{Object} & 
\colhead{observed $\tau_{3.1}$} & 
\colhead{observed $\tau_{3.4}$}  
}
\startdata
UGC 5101 & $>$0.8 & 0.65 \\
IRAS 00188$-$0856 & $>$1.6 & $<$0.2  \\
IRAS 03250+1606   & $>$0.6 & $<$0.1  \\
IRAS 16487+5447   & $>$0.8 & $<$0.35 \\
IRAS 21329$-$2346 & $>$0.9 & $<$0.2  \\
\enddata
\end{deluxetable}

\clearpage

\begin{figure}
\plottwo{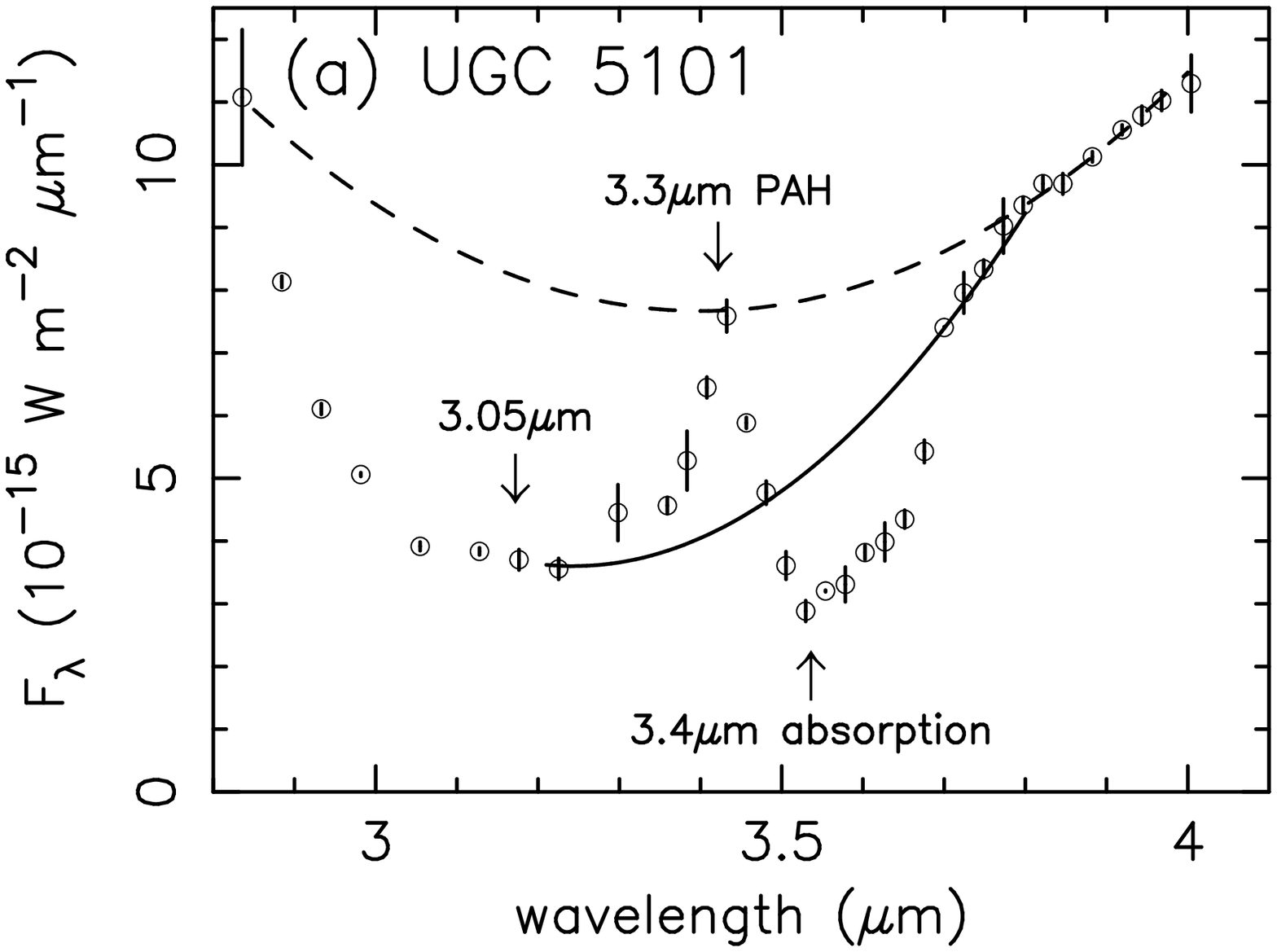}{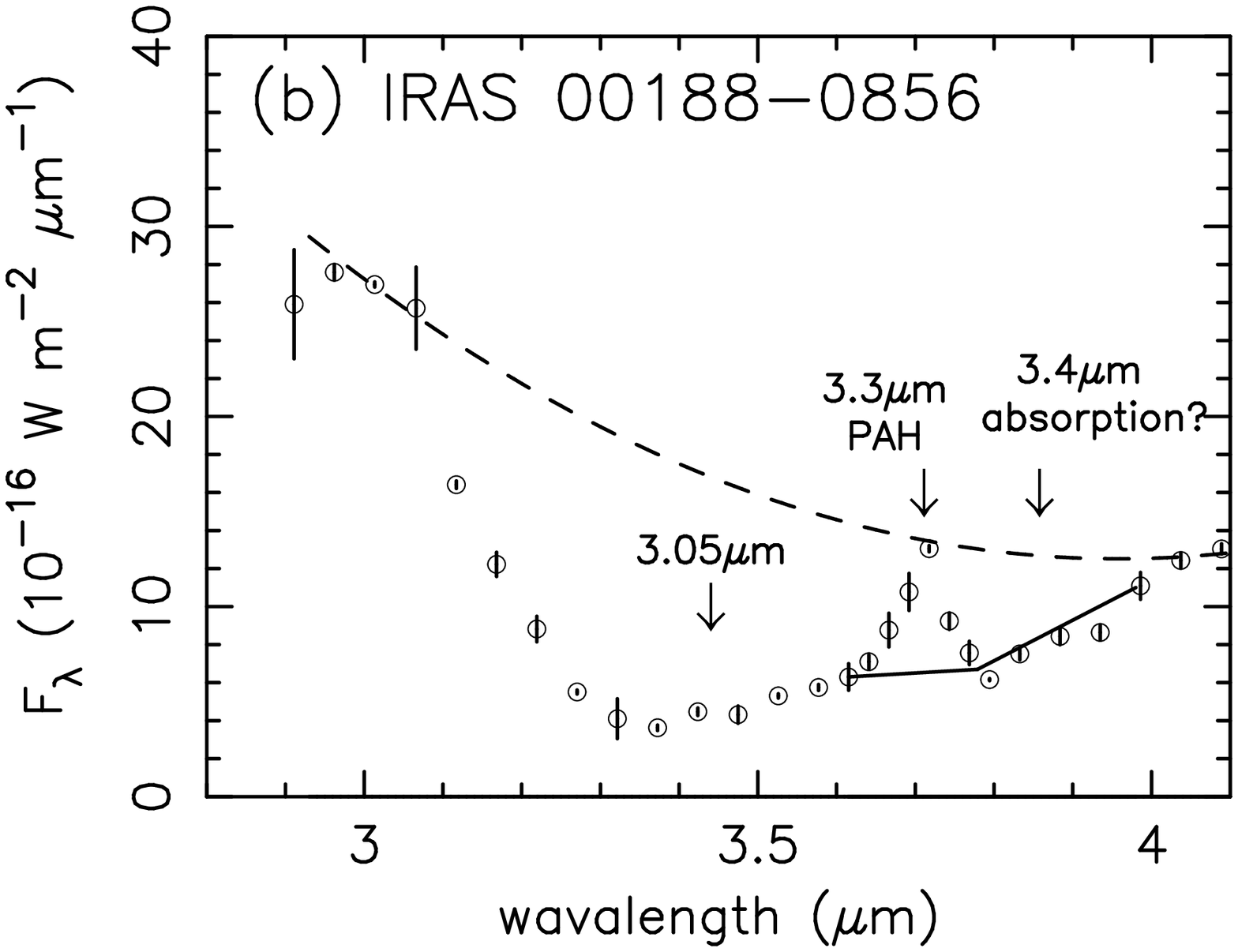}
\plotone{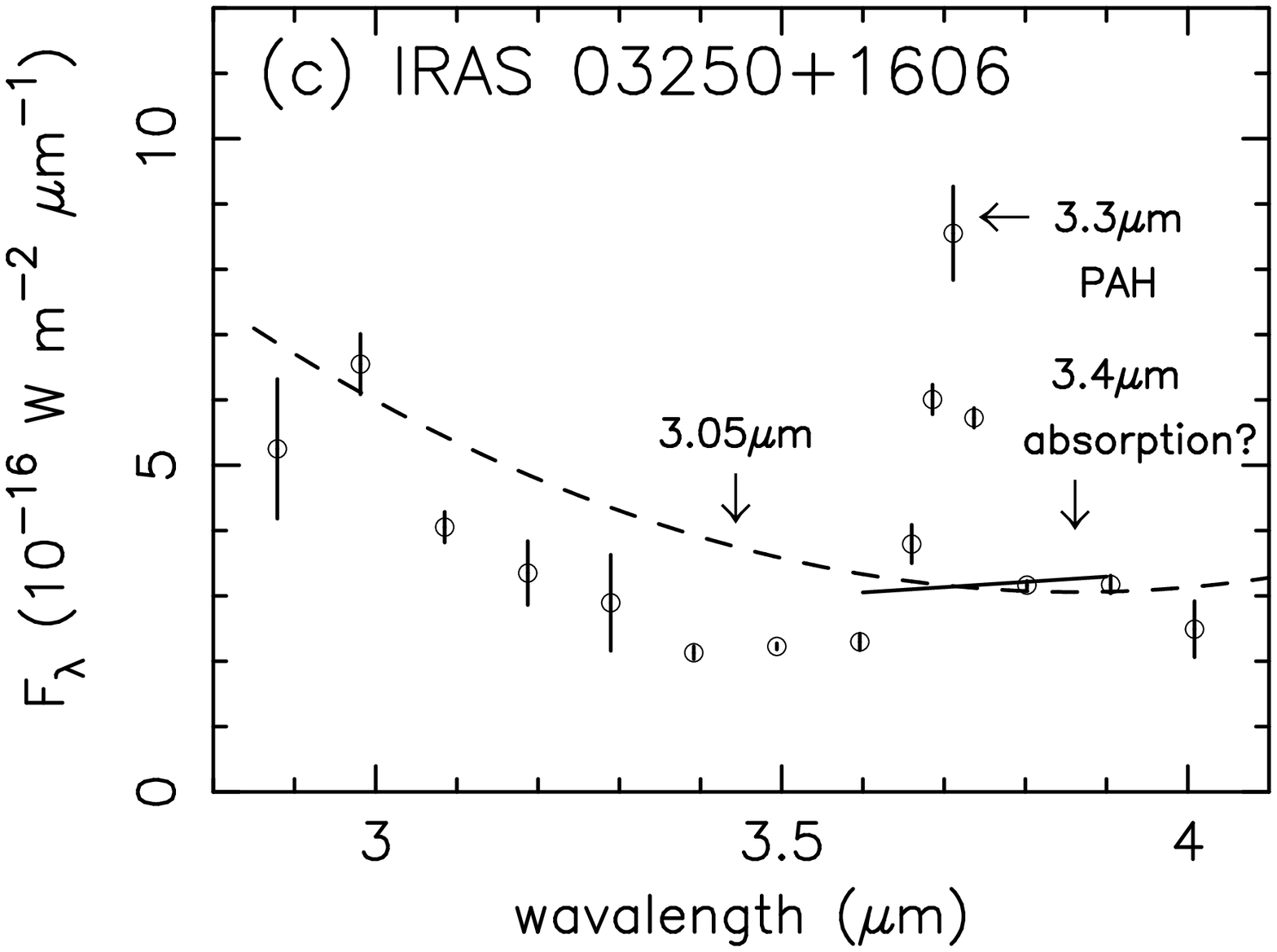}\plotone{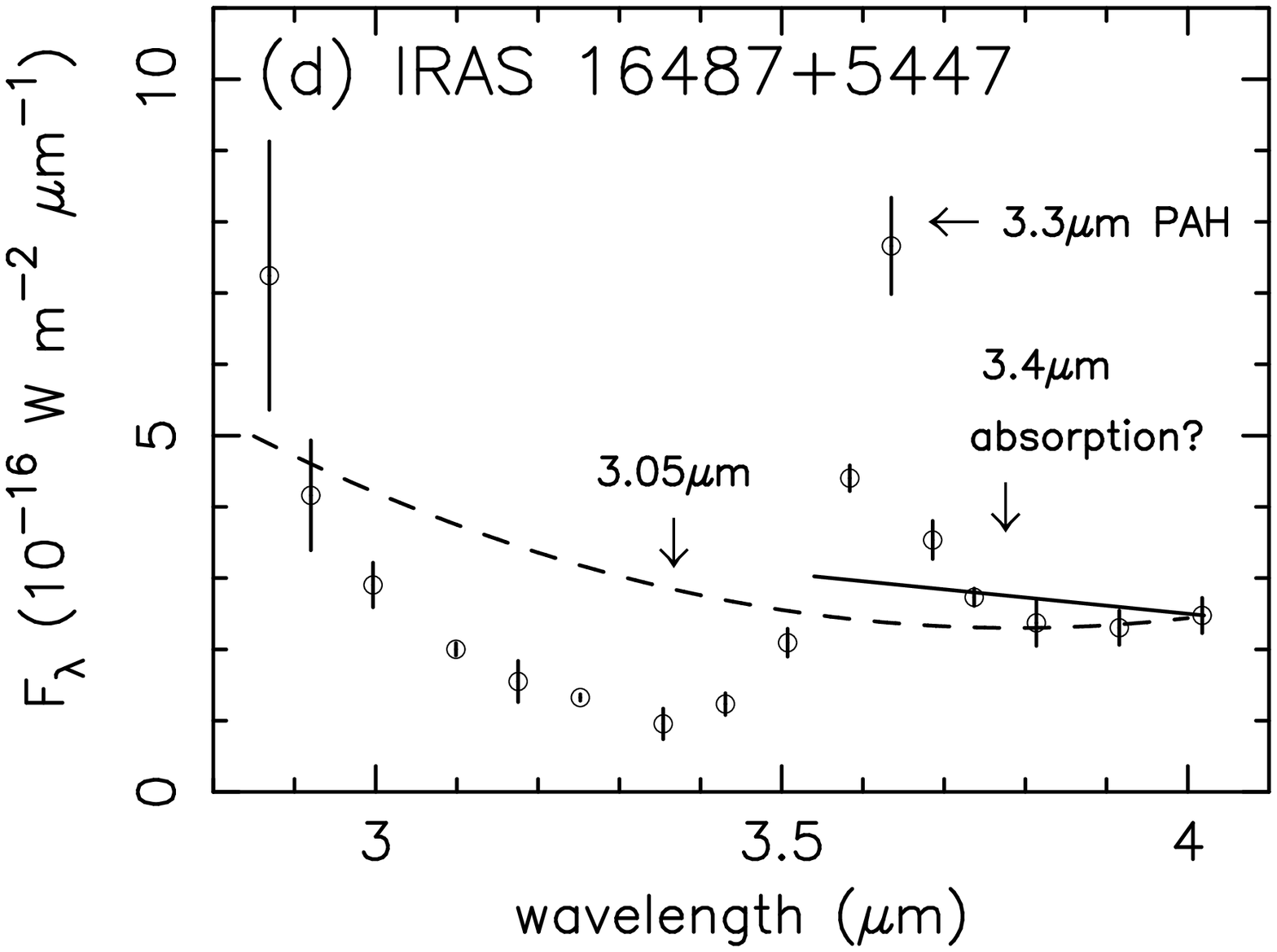}
\end{figure}
\clearpage
\begin{figure}
\plottwo{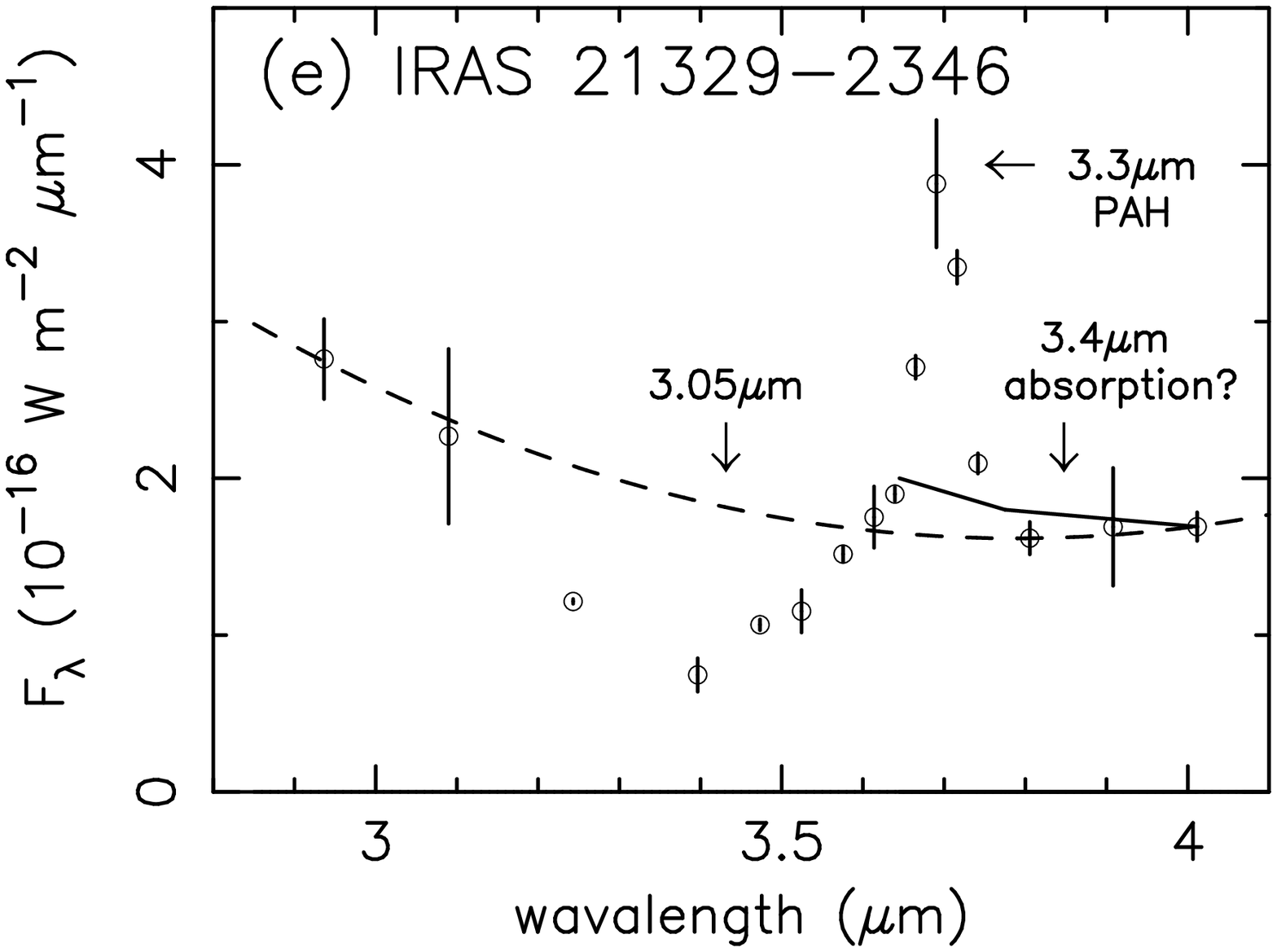}{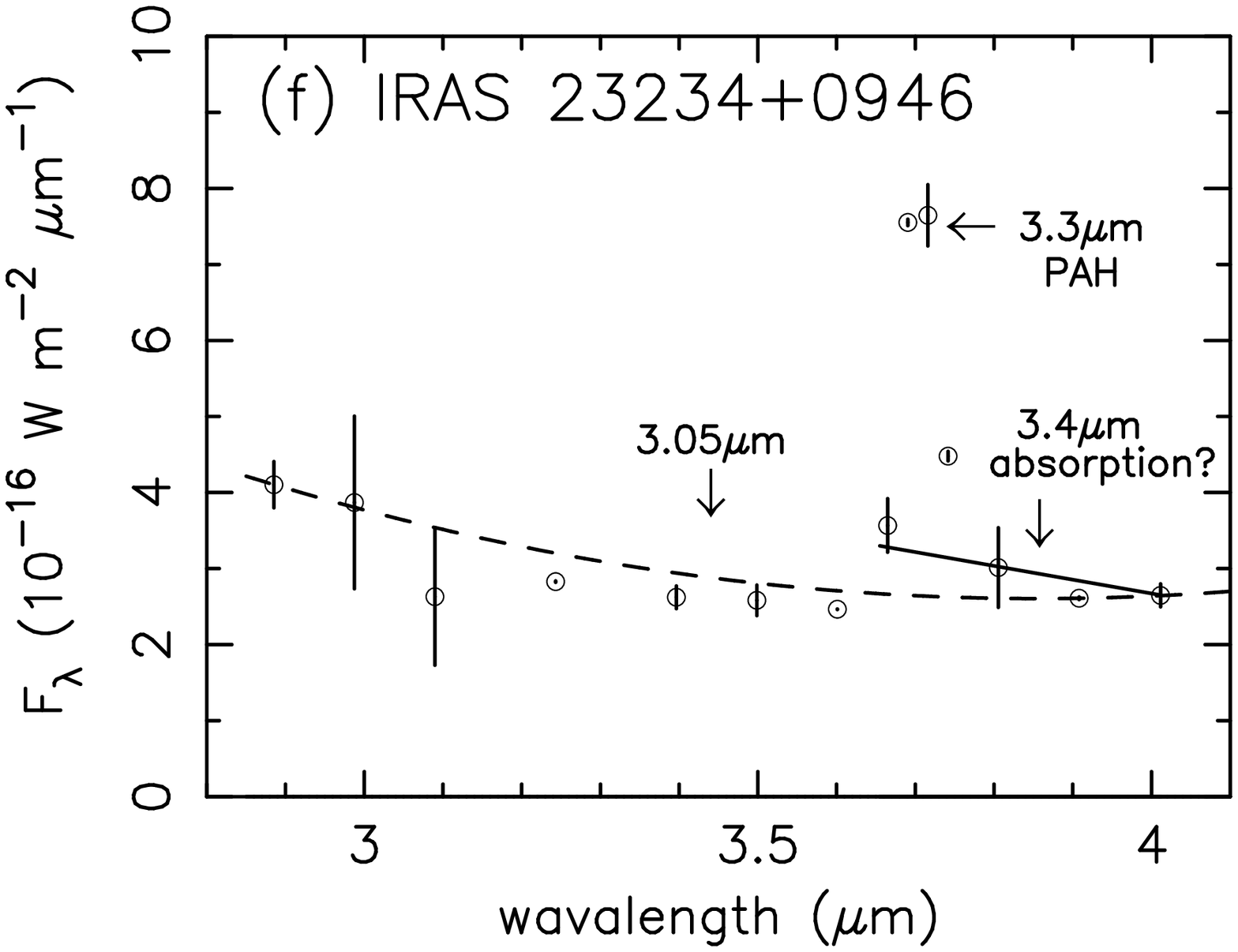}
\plotone{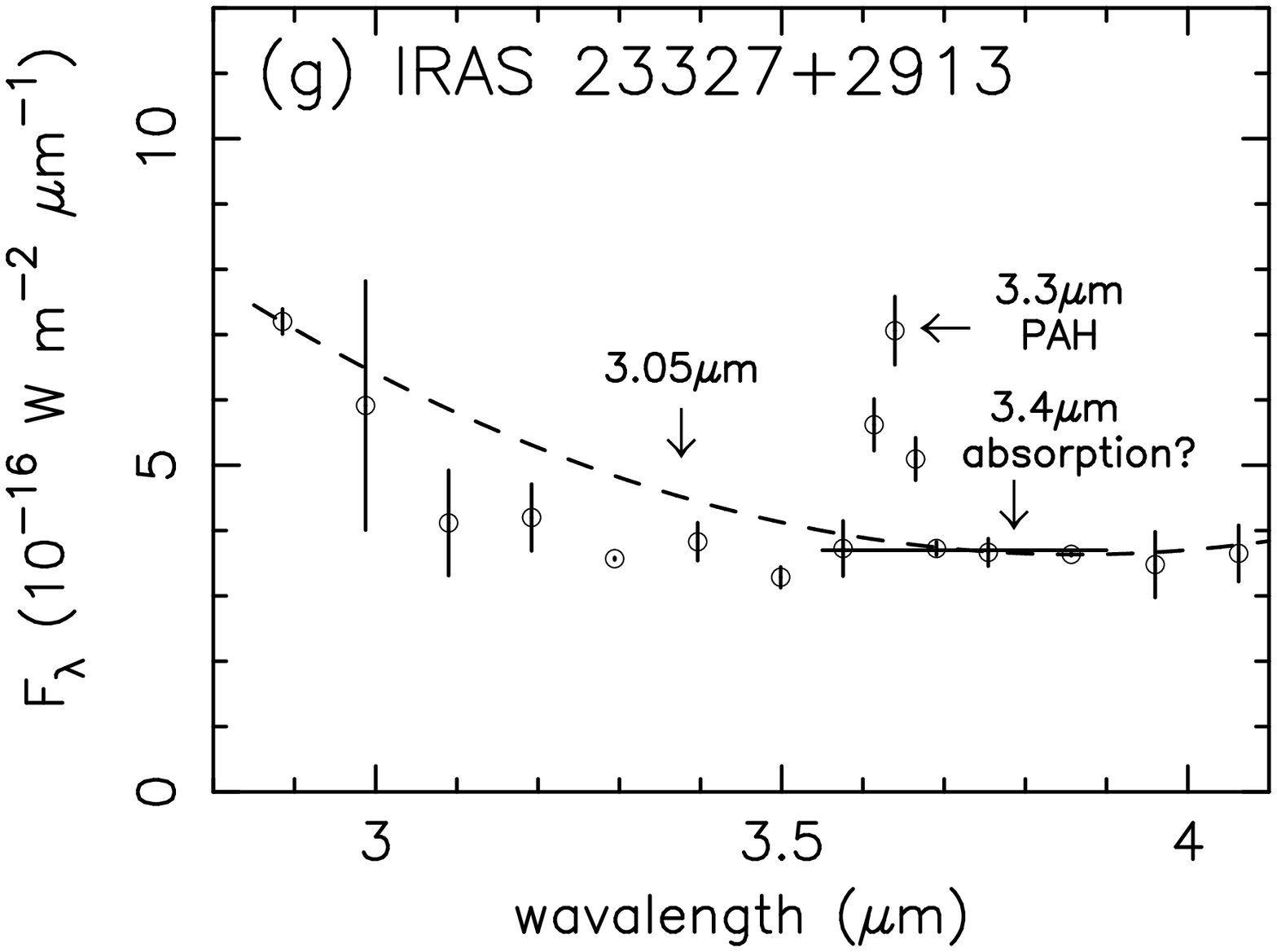}
\caption{
2.8--4.1 $\mu$m spectra of the seven ULIRGs. 
The abscissa and ordinate are the observed wavelength and
the flux in $F_{\lambda}$, respectively.
The dashed concave lines are the continuum levels used to estimate
conservative lower limits for the optical depths of the 3.1 $\mu$m H$_{2}$O
ice absorption. The solid lines are the continuum levels used to estimate
the values of the 3.3 $\mu$m PAH emission and the 3.4 $\mu$m carbonaceous
dust absorption.  
\label{fig2}}
\end{figure}

\begin{figure}
\plottwo{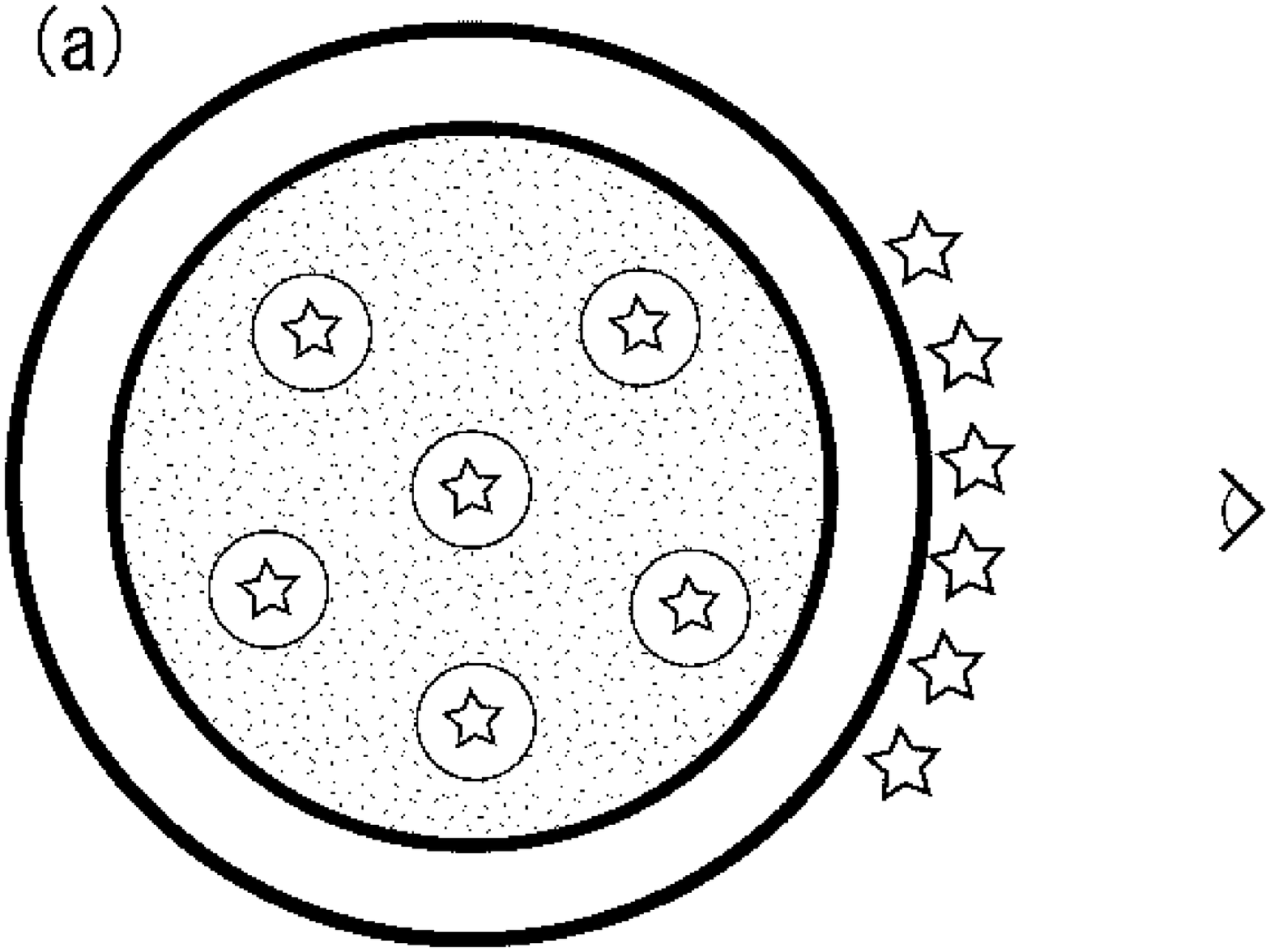}{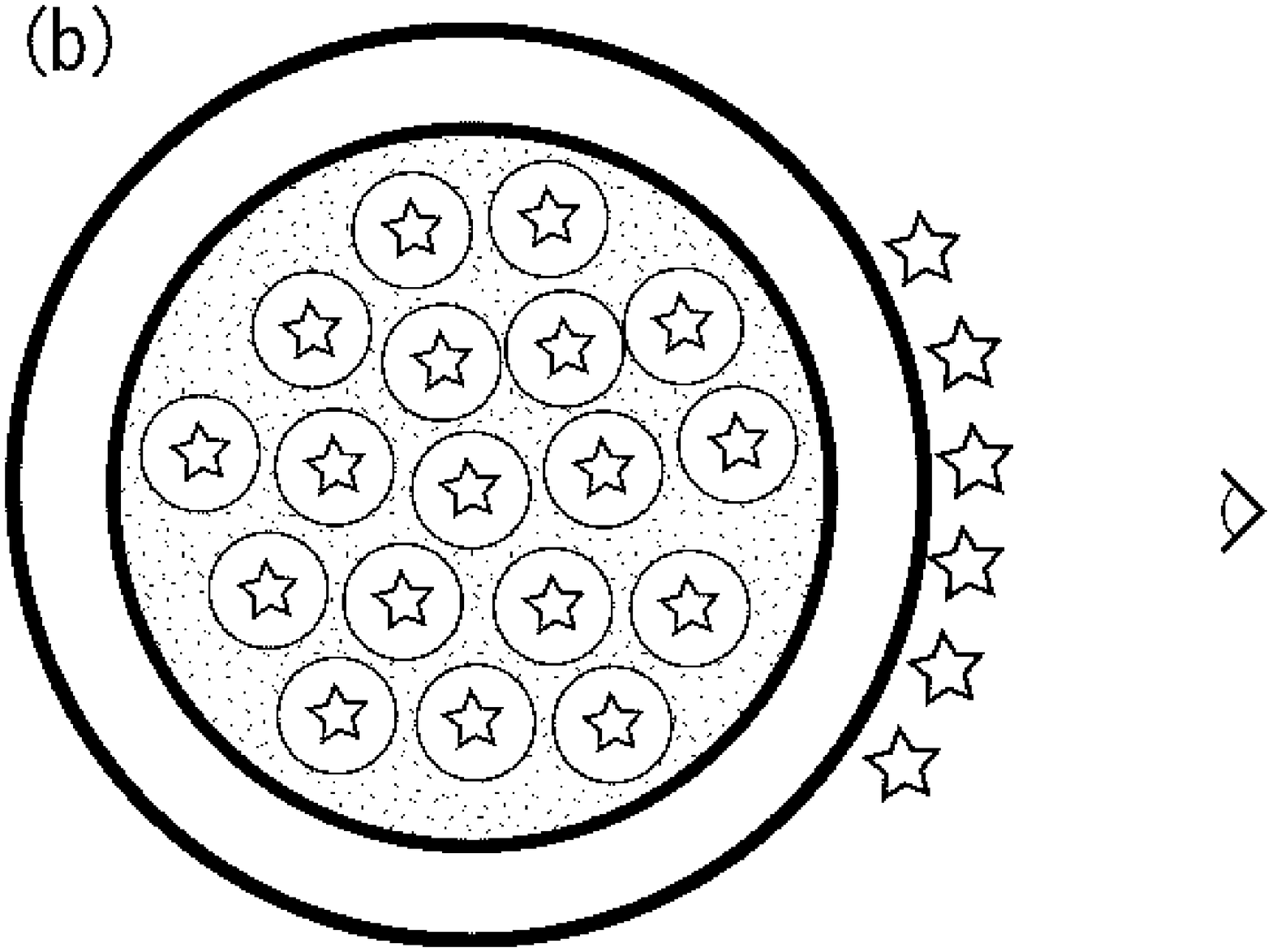}
\plotone{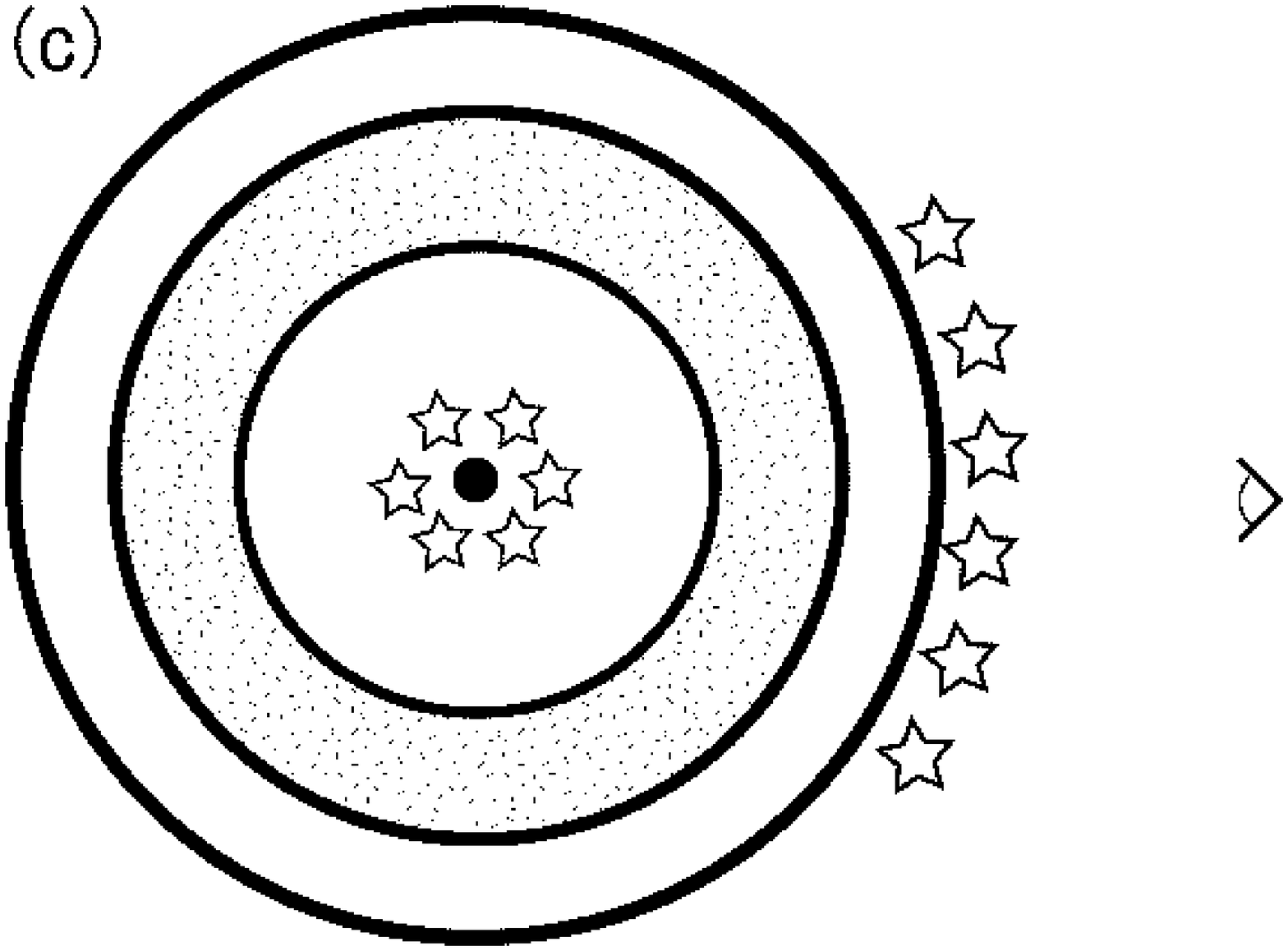}
\caption{
Possible geometries of the energy sources, molecular gas, and
dust.  In all cases, the dotted areas indicate the regions where an
ice mantle can survive (that is, dust is covered with an ice mantle).  
The star symbols represent starbursts.  
The stars outside the outermost circles represent weakly obscured
starbursts; only the side near the observer is shown.  {\it (a)}:
Starburst activity is homogeneously mixed with dust and molecular
gas.  The fraction of active star-forming regions relative to quiescent
molecular gas is modest.  {\it (b)}: The same geometry as {\it (a)}, but
active star-forming regions occupy the majority of the
volume of molecular gas.  {\it (c)}: The buried energy sources are
more centrally concentrated than the dust and molecular gas. The black dot
represents an AGN.  This is the most plausible geometry to account for
the observed 2.8--4.1 $\mu$m spectral shapes of the five ULIRGs with strong
3.1 $\mu$m H$_{2}$O ice absorption (see the text).
\label{fig3}}
\end{figure}

\end{document}